\documentclass[12pt,letterpaper]{article}
\usepackage[utf8]{inputenc}
\usepackage[T1]{fontenc}
\usepackage{amsmath}
\usepackage{amsfonts}
\usepackage{amssymb}
\usepackage{graphicx}
\usepackage{natbib}
\usepackage[includeheadfoot,left=1.in,right=1.in,top=1.in,bottom=1.in,bindingoffset=0.in,nohead]{geometry}
\usepackage[onehalfspacing]{setspace}
\usepackage[pdfborder={0 0 0}]{hyperref}
\usepackage{xcolor}

\title{On the Financing of Climate Change Adaptation\\	in Developing Countries}
\author{Francis X. Diebold \\ University of Pennsylvania}
\date{October 2, 2022}

\begin{document}

\maketitle
\thispagestyle{empty}

\bigskip
\bigskip
\bigskip

\noindent \textbf{Abstract}:  I offer reflections on adaptation to climate change, with emphasis on developing areas.

\bigskip
\bigskip
\bigskip

\begin{spacing}{1}

\noindent \textbf{Acknowledgments:}  Thanks to Aaron Mora, Glenn Rudebusch and participants at the Wharton Climate Center / Perry World House Global Climate Finance Workshop.  The usual disclaimer applies.

\bigskip
\bigskip
\bigskip


\bigskip

\noindent \textbf{Contact:} {fdiebold@sas.upenn.edu}	
	
	\end{spacing}

\clearpage

\setcounter{page}{1}
\thispagestyle{empty}

Let’s start with emissions mitigation rather than adaptation, and let’s first consider mitigation in the developed world.  Developed-world mitigation efforts may fail significantly, for a variety of reasons, ranging from never-ending arrivals of seemingly “more important” situations requiring immediate attention and arguably requiring the postponing mitigation efforts, to the ever-present and powerful incentive to free-ride on others. Indeed developed-world mitigation is already failing significantly (e.g., \cite{unep2021}), a reality that underscores the importance of developed-world adaptation, in addition to mitigation, strategies.

Now move to emissions mitigation in the developing world.  Much of developing-world mitigation may fail not just significantly, but quite spectacularly as mitigation efforts are overwhelmed by the push for higher living standards and further population growth (e.g., \cite{cui2020}).  The rapidly-developing world will require substantial amounts of new energy, and inexpensive fossil fuels will remain highly attractive there\footnote{Moreover, the developing world understandably tends to frown on the rich for now condemning the burning of fossil fuels, given that the rich got rich precisely by burning fossil fuels.}. And in the stagnating part of the developing world, mitigation is a moot issue, as there are few emissions to mitigate.  The entire continent of Africa, for example, contributes little to global emissions.

The bottom line as regards mitigation in the developing world:  Mitigation is likely to be sidelined in the rapidly-developing part of the developing world, and is not of much relevance in the stagnating part of the developing world.  But simultaneously, both parts of the developing world will be forced to adapt.

So adaptation is now emerging as crucially important for both the developed and developing worlds.  Unfortunately, however, climate economics has focused almost exclusively on mitigation issues, such as what Pigouvian tax would price carbon correctly, or whether capital markets penalize brown firms sufficiently such that they are incented to go green, and related, how to define and credibly measure “brown” or “green”.  But carbon-tax proposals are increasingly dead on arrival, and brown-firm capital market penalties alone won’t solve our problems.

Interestingly and fortunately given its increasing role, adaptation has several attractive intrinsic features (e.g., \cite{koonin2021}), some of which are particularly relevant for the developing world:  (1) Adaptation is agnostic.  We don’t need to know and target the source of the climate change.  Whatever the cause, it’s happening, and we can attempt to adapt.  (2) Adaptation isn’t just about policy.  It is aided in significant part by the natural actions of private agents acting in their own self-interest, and it has been working successfully for millennia.  (3) Adaptation project spending, typically for things visible and conveying potential benefits “here and now”, may be more politically palatable than mitigation spending that provides potential benefits to distant future generations.  (4) Adaptation is local geographically.  Different areas have different exposures to different climate risks, and adaptation efforts can themselves be adapted to those differences.  (5) Adaptation is local not only geographically, but also temporally.  It can be scaled up or down as climate risks evolve and climate events unfold.

There are two key, and very different, aspects of climate change that adaptation must confront, one chronic and one acute.  The chronic aspect is associated with gradually-evolving central tendencies.  Front and center is of course average temperature (and closely related, sea level), which is slowly increasing, with concomitant slowly-evolving longer-run issues of increased mortality and illness, reduced labor and agricultural productivity, long draughts, forced migration, conflict, etc.

The acute aspect of climate change that adaptation must confront is not about the central tendencies of climate indicator distributions, but rather about their extreme tails, in particular the frequency and severity of “tail events” like heatwaves, floods, wildfires, and hurricanes.  And even if central tendencies typically evolve linearly and gradually, they can occasionally shift non-linearly and abruptly if “tipping points” are reached, so that chronic aspects of the evolution of central tendencies can themselves feed importantly into acute tail events.

Several developing-world adaptation questions arise.  First, what tools are needed?  One key need is for better data to help with assessing adaptation costs and benefits.  In particular, the developing world needs sustained availability of reliable climate data and damage estimates at a highly-disaggregated level, as well as associated reliable and highly-disaggregated climate model projections, not platitudes like “the sea level will continue to rise”.  The Climate Impact Lab (\href{impactlab.org}{impactlab.org}) is making important advances in that direction (e.g., \cite{carleton2022}).

Second, what should appear in developing-country adaptation budgets?  They should be sharply targeted, including climate-resistant infrastructure; heat-resistant, drought-resistant, and flood-resistant seeds and varietals; improved risk-management and risk-transfer tools, ranging from better crop diversification to climate derivatives and insurance; improved climate event early-warning systems using real-time data relayed from drones and satellites; and not at all least, services for displaced people.  It will be crucial, moreover, to add adaptation expenditures without removing other well-targeted items from existing budgets, like vaccines, antibiotics, fertilizer, and medical personnel.

Third, how can developing-country adaptation be financed?  There are many issues: bureaucratic, political, and fiscal.  Much of local project-based adaptation has a public-good aspect, rendering purely private-market solutions suboptimal and creating a role for governmental oversight.  And quite apart from the public-good aspect, the scale of adaptation infrastructure projects (e.g., sea walls) promotes natural monopolies, which again create a role for governmental oversight.

“Adaptation as public expenditure” would therefore seem to involve little that is new in terms of financing, as the government budget constraint remains inescapable:  countries can inflate, tax, borrow, or accept aid.  The problem is that while rich countries can often afford to pay the “adaptation tax”, developing countries often can’t.  Hence aid may be an important part of developing-country adaptation finance, at least in the short run, whether directly via foreign aid, or indirectly via subsidized loans from development banks or specialized vehicles like the Green Climate Fund (GCF)\footnote{Whether this aid should be considered a form of reparations for past carbon pollution remains a contentious issue.}. 

In the medium and long runs, however, the success record of aid is mixed at best (\cite{easterly2017}).  Among other things, it creates undesirable incentives, and it may be appropriated by corrupt bureaucrats.  Hence, in the longer run, successful adaptation will benefit from non-aid financing.  

Venture capital finance of new adaptation technologies may prove especially useful.  Markets for venture capital, like all markets, function best in environments of capitalism, democracy, and law (that is, well-defined and well-enforced property rights), which are often lacking in developing countries.  But an important twist arises:  Certain adaptation technologies may effectively be financed and implemented in developed countries, but with all countries – developed, rapidly-developing, and stagnating – nevertheless benefiting equally.  

A leading example is carbon capture and storage (CCS), which blends aspects of both adaptation and mitigation.  CO2 mixes very quickly in the earth’s atmosphere, so that an emissions burst in Houston, for example, will quickly contribute to atmospheric CO2 concentration globally and equally.  Conversely, CO2 capture and storage in Houston will reduce atmospheric CO2 concentration globally and equally.

The massive positive externality associated with successful CCS is certainly appealing, but there remains the issue of large-scale feasibility.  The news there is quite good: CCS technology is improving, and attractive high-capacity geological storage locations are available (e.g., \cite{aydin2010}; \cite{ball2022}).  Leading candidates include deep saline aquifers and depleted fossil-fuel reservoirs.

Investor-led venture capital funds like Breakthrough Energy Ventures (\href{breakthroughenergy.org}{breakthroughenergy.org}) may play a major role in financing CCS and other emerging technologies, which hold promise not only for reducing the annual global carbon flow, but also for reducing the global carbon stock, which is what ultimately drives warming\footnote{See also \cite{gates2021}.}.  

\bibliographystyle{Diebold}

\bibliography{Bibliography}
	
\end{document}